\documentclass{emulateapj}

\slugcomment{to appear in ApJ}

\shorttitle{X-Ray Monitoring of ULXs}
\shortauthors{Kaaret \& Feng}

\begin{document}

\title{X-Ray Monitoring of Ultraluminous X-Ray Sources}

\author{Philip Kaaret\altaffilmark{1} and Hua Feng\altaffilmark{2}}

\altaffiltext{1}{Department of Physics and Astronomy, University of
Iowa,  Van Allen Hall, Iowa City, IA 52242, USA}

\altaffiltext{2}{Department of Engineering Physics and Center for
Astrophysics, Tsinghua University, Beijing 100084, China}

\begin{abstract}

X-ray monitoring observations were performed with the {\it Swift}
observatory of the ultraluminous X-ray sources Holmberg IX X-1, NGC 5408
X-1, and NGC 4395 X-2 and also of the nuclear X-ray source in NGC 4395. 
Holmberg IX X-1 remains in the hard X-ray spectral state as its flux
varies by a factor of 7 up to an (isotropic) luminosity of $2.8 \times
10^{40} \rm \, erg \, s^{-1}$.  This behavior may suggest an unusually
massive compact object.  We find excess power at periods near 60~days
and 28~days in the X-ray emission from Holmberg IX X-1. Additional
monitoring is required to test the significance of these signals.  NGC
5408 X-1 and NGC 4395 X-2 appear to remain in the soft spectral state
found by Chandra and XMM with little variation in spectral hardness even
as the luminosity changes by a factor of 9.  We found an outburst from
the nuclear source in NGC 4395 reaching an X-ray luminosity of $9 \times
10^{40} \rm \, erg \, s^{-1}$, several times higher than any previously
reported.

\end{abstract}


\section{Introduction}

Ultraluminous X-ray sources (ULXs) were originally identified as
potential intermediate-mass black holes (IMBHs) on the basis of the high
luminosities inferred assuming isotropic emission of X-rays
\citep{Colbert99,Makishima00,Kaaret01}.  If ULXs are, indeed, radiating
isotropically below the Eddington luminosity, then the inferred masses
are, in some cases, greater than $500 M_{\odot}$.   This is larger than
the maximum compact object mass which can be formed in the collapse of a
single star with metallicity $Z \gtrsim 10^{-3} Z_{\odot}$
\citep{Bromm03}, and requires a different mechanism for formation.  The
existence of IMBHs would be of interest for subjects ranging from the
formation of galaxies and their supermassive black holes to the
generation of gravitational waves.  However, the physical nature of ULXs
is not well understood.  The X-rays may be mechanically or
relativistically beamed in which case IMBHs are not required. 

New information is needed to understand the ULXs.  Knowledge of the
patterns of evolution of the X-ray emission from Galactic black hole
X-ray binaries has been key to understanding their physical nature
\citep{Remillard06,Belloni05}.  Detailed study of spectral evolution was
key to determining the physical nature of the spectral components
contributing to the emitted X-ray flux.

Previously, it has been possible to obtained light curves with more than
about a dozen points only for the brightest ULX located in M82
\citep{Kaaret06sci,Kaaret06,Kaaret07}.  The {\it Swift} observatory
\citep{Gehrels04} has a capability, unique among focusing X-ray
telescopes, to perform multiple observations of a given target with
flexible scheduling.  While the effective area of the Swift X-Ray
Telescope (XRT) is limited in comparison with the larger X-ray
observatories, the brighter ULXs have sufficiently high fluxes to
provide reasonable numbers of counts ($\sim 200$) in modest individual
observations (less than 2~ks).  Thus, Swift provides the unique means to
extend our knowledge of ULXs by probing their patterns of spectral
evolution with dense temporal coverage.

Here, we report on X-ray monitoring observations performed with the {\it
Swift} observatory of the ultraluminous X-ray sources Holmberg IX X-1,
NGC 5408 X-1, and NGC 4395 X-2 and also of the nuclear X-ray source NGC
4395 X-1.  The targets, observations, and data reduction are described
in \S 2. The results are discussed in \S 3.

\section{Targets, Observations, and Data Analysis}

Holmberg IX X-1 (= NGC 3031 X-9) has an average X-ray luminosity near
$10^{40} \rm \, erg \, s^{-1}$ and is surrounded by a large and
energetic optical nebula \citep{Grise06}.  This ULX is one of the best
on which to perform long term monitoring with Swift because it is
brightest ULX with no contaminating sources within the angular
resolution of the Swift XRT.  The source shows large flux variations
across the observations obtained to date by various X-ray instruments
\citep{LaParola01}.   Based on XMM-Newton spectra, we estimated a
typical count rate for the Swift XRT of 0.14~c/s when retaining events
with grades 0-12 in photon counting mode.  Thus, measurement of the
source intensity and a hardness ratio are possible with observations of
around 1400~s. We obtained 72 observations of Holmberg IX X-1 under
Swift program 90008 (PI Kaaret).  In addition, we analyzed one
observation obtained under program 25952, 9 from program 35335, and 25
from program 90079.

We also searched the Swift archive for series of multiple observations
of other relatively bright ULXs.  We found 78 observations of the ULX
NGC 5408 X-1 under program 90041 and 24 under program 90218 (both PI
Strohmayer).  This ULX has exhibited quasiperiodic oscillations at
relatively low frequencies \citep{Strohmayer07} and is surrounded by a
powerful radio nebula \citep{Kaaret03,Soria06,Lang07} and a photoionized
optical nebula \citep{Kaaret09}.  NGC 5408 X-1 is one of the best
intermediate mass black hole candidates amongst the ULXs.  However, NGC
5408 X-1 is dimmer than Holmberg IX X-1.

We also found 59 observations of the nearby active galaxy NGC 4395
obtained under program 90053 (PI Uttley).   NGC 4395 contains an X-ray
source at a position near RA = 12h26m02s, DEC 33d31'34" (J2000), NGC
4395 X-2 in \citet{Lira00} and source E in \citet{Moran99}, that during
the ROSAT-era appeared brighter than nuclear X-ray source, reaching a
luminosity near $3 \times 10^{39} \rm \, erg \, s^{-1}$.  We examined
this source.  We also examined the nuclear X-ray source in NGC 4395, NGC
4395 X-1 in \citet{Lira00} and source A in \citet{Moran99}, for
comparison with the ULXs.

We retrieved level 2 event files with observations of these targets made
in photon counting and pointed mode.  These event files have data
screening already applied as described in the XRT User's Guide.  Each
observation is divided up into one or more good time intervals (GTIs).
We analyzed each GTIs separately, retaining only those GTIs with
durations of 100~s or longer.  We extracted source counts from a region
with a radius of $25\arcsec$ and background counts from an annulus with
an inner radius of $50\arcsec$ and an outer radius of $100\arcsec$. 
This extraction radius captures 82\% of the total photon flux
\citep{Moretti06} and we correct for this when calculating source
fluxes.  There are bad pixels in the XRT CCD that can lead to a loss of
photons.  For each GTI, we calculated the displacement of the center of
gravity of the recorded photons from the nearest bad pixel and rejected
the GTI if the displacement was less than $12.5\arcsec$.  This removes
GTIs that have one or more bad pixels near the core of the source point
spread function.  About 20\% of the GTI were rejected due to bad pixels.
Using the measured point spread function \citep{Moretti06} and making
the conservative assumption that the nearest bad pixel is at the edge of
a half-plane of bad pixels, we estimate that the maximum loss of counts
for the unrejected GTIs is less than 6\%.  We performed our analysis
using the Pulse Invariant (PI) data which are corrected for gain
variations with time, temperature, and charge transfer inefficiency. 
For each observation interval, we calculated the net count rate in the
0.3-8~keV band and in the 1-8~keV band.  We calculated a hardness ratio
equal to the rate in the hard band (1-8~keV) divided by the rate in the
full band (0.3-8~keV).

\begin{figure*}[tb]
\centerline{\includegraphics[angle=0,width=6.5in]{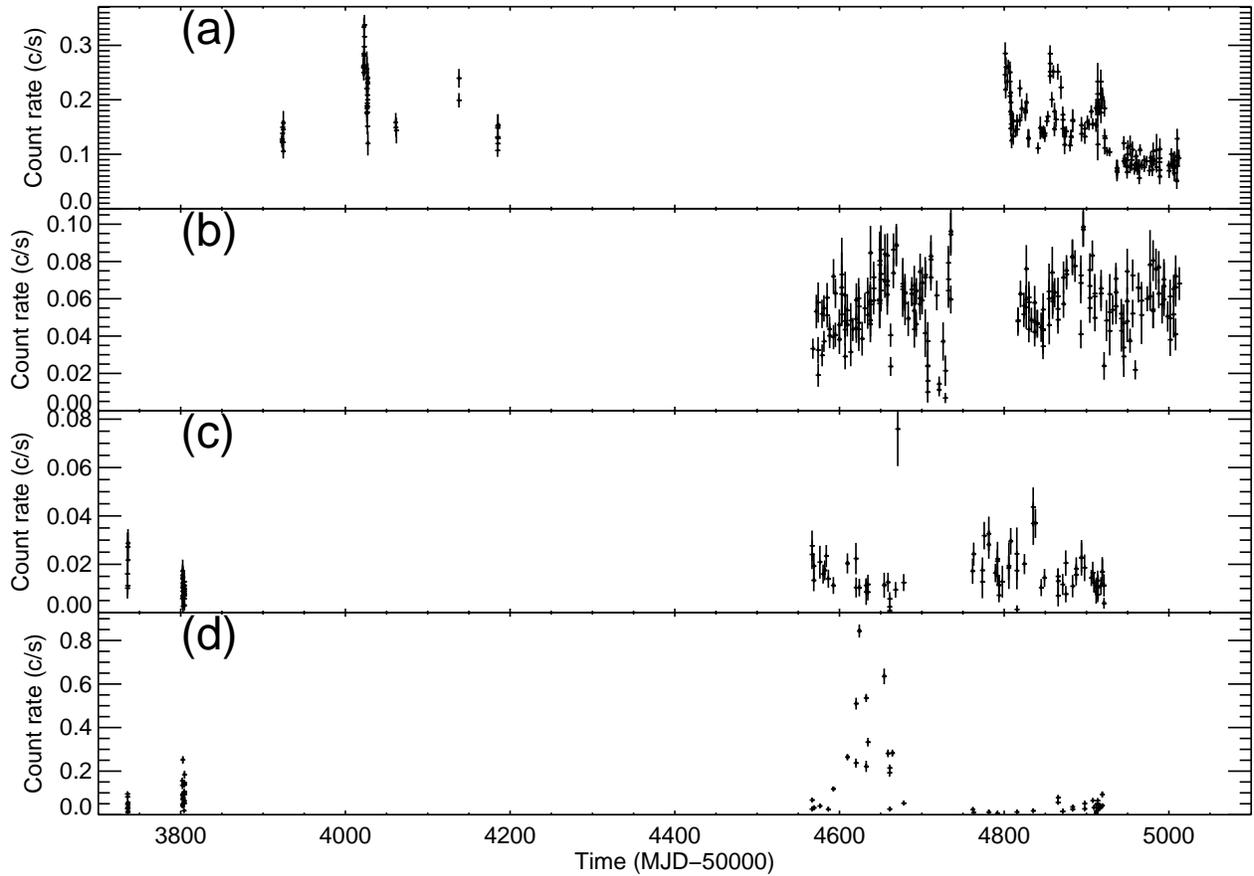}} \caption{X-ray
light curves in the 0.3-8~keV band for: (a) Holmberg IX X-1, (b) NGC
5408 X-1, (c) the ULX NGC 4395 X-2, (d) the nuclear source NGC 4395
X-1.}  \label{lc}  \end{figure*}

\begin{figure}[tb]
\centerline{\includegraphics[width=3.25in]{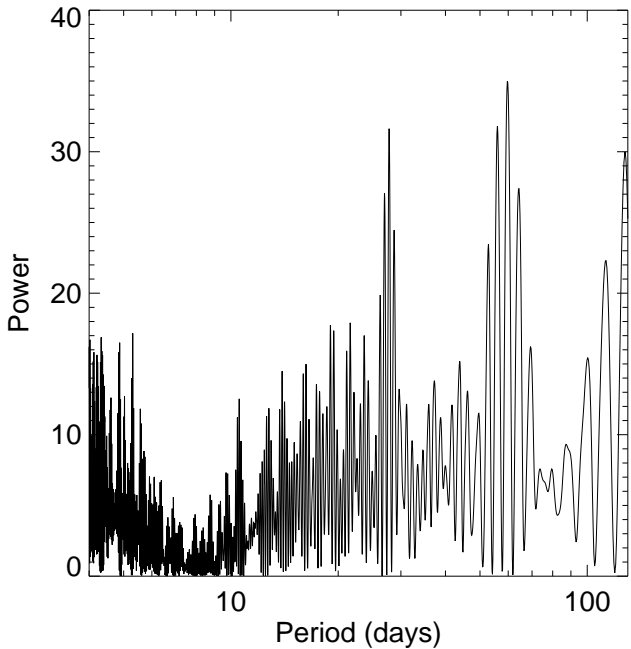}}
\caption{Periodogram calculated from the X-ray flux measurements of
Holmberg IX X-1.  Peak are apparent at a periods of 28.8 and 64.5~days.}
\label{hoix_period} \end{figure}

\begin{figure}[tb] \centerline{\includegraphics[width=3.25in]{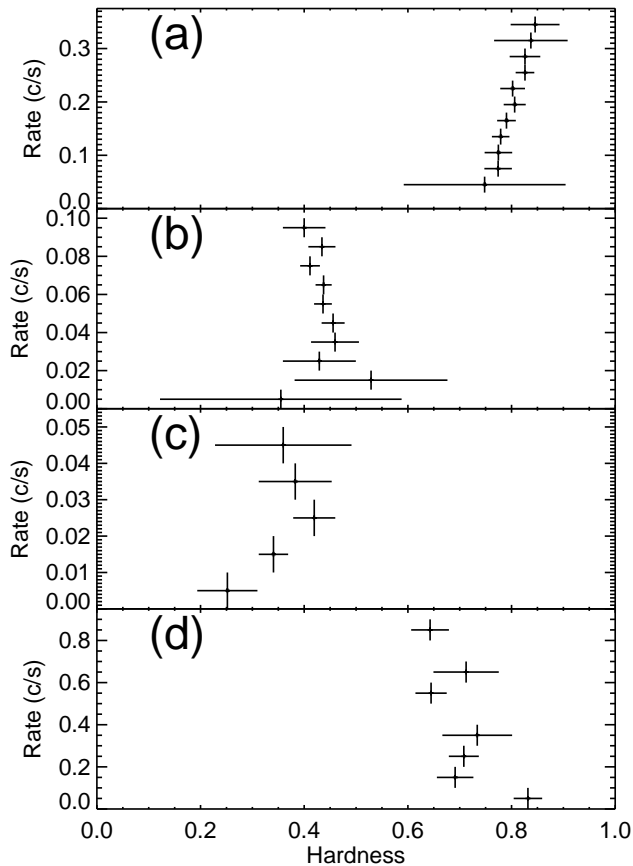}}
\caption{Hardness/intensity diagrams for the four sources: (a) Holmberg
IX X-1, (b) NGC 5408 X-1, (c) the ULX NGC 4395 X-2, (d) the nuclear
source NGC 4395 X-1.} \label{hi} \end{figure}

\begin{deluxetable}{lccc} \tabletypesize{\scriptsize}
\tablecaption{Spectral parameters of ULXs \label{hardtable}}
\tablewidth{0pt} 
\tablehead{ 
           & \colhead{Holm IX X-1}
           & \colhead{N5408 X-1}
           & \colhead{N4395 X-2} }
\startdata 
Hardness         & 0.80$\pm$0.01 & 0.43$\pm$0.01 & 0.35$\pm$0.02 \\
$\chi^2$/DoF     & 7.9/10        & 4.0/9         & 6.1/4         \\
$n_H$ (cm$^{-2}$)
                 & $2.0 \times 10^{20}$
                                 & $1.4 \times 10^{20}$
		                                 & $2.4 \times 10^{21}$ \\
$\Gamma$         & 1.88          & 3.06          & 4.12          \\
Hardness         & 0.79          & 0.47          & 0.39          \\
\enddata 
\tablecomments{The table includes for each source: the average
X-ray hardness measured with Swift, the column density and photon index
measured from fits to XMM-Newton data of an absorbed power-law model,
and the expected Swift hardness calculated using the XMM-Newton spectral
fits.} \end{deluxetable}

\section{Results}

The X-ray light curves of the four targets in the 0.3-8~keV band are
shown in Fig.~\ref{lc}.  Significant variability is seen in all of the
light curves.  We calculated periodograms for the data shown in each
light curve using the method of \citet{Horne86} with the power
normalized by the total variance of the data for periods in the range of
4 to 130 days.  The maximum power is 18.1 near a period of 115 days for
NGC 5408 X-1 and 12.3 at a period near 28 days for NGC 4395 X-2; neither
is a significant signal.  The periodogram for Holmberg IX X-1 is shown
in Fig.~\ref{hoix_period}.  There are two peaks: one with a power of
35.0 at a period near 60~days and one with a power of 31.6 at a period
near 28~days.  The fine structure of the peaks is due to aliasing of
nearby periods due to the large gap in time between the different epochs
of Swift observations. We note that a significant part of the power
comes from the flare near MJD 54020 in observations taken before the
start of the monitoring.  Excluding the data from MJD 54000 to 54100,
the strongest signal has a power of 23.1 at a period near 56~days. 
Thus, we do not interpret the signal as a true periodicity at this
time.  However, the observed signal is strong motivation to continue
monitoring of Holmberg IX X-1 with {\it Swift}.

Hardness/intensity diagrams for the four targets are shown in
Fig.~\ref{hi}.  The intensity is the count rate in the 0.3-8~keV band
and the hardness is the ratio of the rate in the hard band (1-8~keV)
divided by the rate in the full band (0.3-8~keV).   The individual good
time intervals were generally too short to adequately constrain the
hardness ratio, particularly for the dimmer sources.   However, the
hardness values for a given source at a given count rate appeared
consistent.  Thus, we chose to bin the data by count rate and calculate
the average hardness for all of the observations within each count rate
bin.  The behavior of all three ULXs is consistent with no significant
spectral evolution as a function of intensity.  The average hardness and
the $\chi^2$ for a model where the hardness is taken as equal to the
average are shown in Table~\ref{hardtable}.  For comparison, we show the
results of a fit of an absorbed power-law model to XMM-Newton data
\citep{Feng05} and the hardness that would be obtained observing such a
spectrum with Swift.  To estimate the sensitivity of the hardness ratio,
we note that changing the photon index for Holmberg IX X-1 from 1.88 to
2.00 changed the hardness from 0.79 to 0.77.  In all cases, the hardness
calculated from the best fit XMM-Newton spectrum is in good agreement
with the average hardness found from the Swift data.  For Holmberg IX
X-1, we separately examined the data with rates between 0.06~c/s and
0.3~c/s.  The average hardness is $0.80 \pm 0.01$ and the goodness of
fit for a constant model is $\chi^2$/DoF = 5.8/7.  Thus, the data remain
consistent with constant hardness even if the points with large errors
are excluded.

\citet{Remillard06} have identified three main spectral/timing states of
accreting stellar-mass black holes: the steep power-law state, the
thermal dominant state, and the hard state.   The hard state is defined
as having a photon index $1.4 < \Gamma < 2.1$.  The Swift data show that
Holmberg IX X-1 remains in the hard state, with $\Gamma$ near 1.9, as
the flux varies over a factor of $\sim$7.  The highest counting rate
observed with Swift corresponds to a flux of $1.8 \times 10^{-11} \rm \,
erg \, cm^{-2} \, s^{-1}$ in the 0.3-10~keV band using the spectral
model in Table~\ref{hardtable} and a luminosity of $2.8 \times 10^{40}
\rm \, erg \, s^{-1}$ at the distance of 3.6~Mpc to Holmberg IX assuming
isotropic emission.  Correcting for absorption increases this luminosity
by 50\%.  

Stellar-mass black hole binaries can stay in the hard state  for
extended periods at luminosities below around $0.05 L_{\rm Edd}$, where
$L_{\rm Edd}$ is the Eddington luminosity.  In particular, GX 339-4 has
been observed to remain in the hard state for intervals of more than one
year \citep{Remillard06,Miyakawa08}.  Thus, the behavior observed from
Holmberg IX X-1 may be similar to that observed from GX 339-4, but with
a higher luminosity threshold for the transition out of the hard state
implying a higher Eddington luminosity.  If so, then an unusually
massive black hole is required for Holmberg IX X-1.  We note that some 
stellar-mass black holes have been observed in the hard state at
luminosities as high as $0.3 L_{\rm Edd}$
\citep{Rodriguez03,Zdziarski04,Yuan07,Miyakawa08}.  However, such high
hard-state luminosities are soon followed by a transition to a softer
spectral state, which is not observed in Holmberg IX X-1.

The behavior of Holmberg IX X-1 is similar to that observed from a
number of other particularly luminous ULXs that remain in the hard state
even at the highest luminosities
\citep{Soria07,Berghea08,Kaaret09m82,Soria09,Feng09}.  These ULXs with
hard spectra at luminosities above $10^{40} \rm \, erg \, s^{-1}$ are a
particularly interesting class of objects and deserve further study.

The spectra of NGC 5408 X-1 and NGC 4395 X-2 are softer than found in
the hard state.  These sources appear to be in the steep power-law
state, defined as a photon index $\Gamma > 2.4$.  The Swift observations
find these sources at luminosities $\lesssim 1 \times 10^{40} \rm \, erg
\, s^{-1}$.  There are several other ULXs that appear in the steep
power-law state at similar luminosities \citep{Feng05}.  The steep
power-law state tends to occur at the highest luminosities seen from
accreting stellar-mass black holes.  These ULXs may represent a high
luminosity extension of the steep power-law state.

NGC 4395 has been referred to as the least luminous type 1 Seyfert
galaxy \citep{Moran99} due to measurements of very low luminosities with
ROSAT.  More recent observations covering a broader energy band suggest
an average luminosity near $9 \times 10^{39} \rm \, erg \, s^{-1}$ in
the 0.5-10~keV band \citep{Moran05}, within the range seen from ULXs.  
The light curve of the nucleus of NGC 4395 shows a strong flare near MJD
54624.  We use a simple absorbed power-law model to convert the Swift
count rates to fluxes.  Following \citet{Moran99}, we adopt an
absorption column density $N_H = 1.6 \times 10^{20} \rm \, cm^{-2}$.  We
use a photon index $\Gamma = 1.5$ which produces a hardness of 0.7, in
reasonable agreement with the measured values.  The true X-ray spectrum
of NGC 4395 X-1 is more complex than this, but this approximation should
sufficient to produce rough flux estimates.  The peak counting rate
observed is 0.84~c/s and corresponds to a flux of $4.5 \times 10^{-11}
\rm \, erg \, cm^{-2} \, s^{-1}$ in the 0.5-10~keV band and a luminosity
of $9.1 \times 10^{40} \rm \, erg \, s^{-1}$ at the distance of
4.1~Mpc.  This is several times higher than any flux previously reported
from NGC 4395 X-1.

The spectral evolution of the nuclear X-ray source in NGC 4395 is
inconsistent with being constant.  The $\chi^2$/DoF for a model of
constant hardness is 27.9/6 corresponding to a probability of occurrence
of $1.0\times 10^{-4}$.  The spectrum appears to harden at the lowest
flux levels observed.  \citet{ONeill06} found a similar trend comparing
two Chandra observations of NGC 4395 and suggested that the cause is
variable absorption.  NGC 4395 is known to exhibit dramatic long-term
spectral variability on time scales of several years.  Such dramatic
variability is not apparent in the Swift data, but this may be because
the Swift data cover only one year.

The monitoring programs described here demonstrate that Swift can be
used to measure the flux and spectral evolution of ULXs and AGN on time
scales of months to years.  For the nuclear source in NGC 4395, the
trend for harder spectra at lower fluxes seen in two Chandra
observations \citep{ONeill06} is confirmed with a sample of more than
100 observations and the highest X-ray luminosity ever seen, $9 \times
10^{40} \rm \, erg \, s^{-1}$, was recorded.  The three ULXs monitored
do not show significant changes in spectral state over months to years. 
Two of the ULXs, NGC 5408 X-1 and NGC 4395 X-1, remain in a soft
spectral state, equivalent to a photon index softer than 2.6, as their
flux varies by factors of $\sim 9$.  The other ULX, Holmberg IX X-1,
remains in a hard spectral state, equivalent to a photon index near 1.9,
as its flux varies by a factor of 7 in observations spread over several
years and with (isotropic) luminosities up to $2.8 \times 10^{40} \rm \,
erg \, s^{-1}$.  This behavior may suggest an unusually massive compact
object.

%\section*{Acknowledgments}

%--------------

\end{document}